\documentclass[fleqn,usenatbib]{mnras}

\usepackage{newtxtext,newtxmath}

\usepackage[T1]{fontenc}

\DeclareRobustCommand{\VAN}[3]{#2}
\let\VANthebibliography\thebibliography
\def\thebibliography{\DeclareRobustCommand{\VAN}[3]{##3}\VANthebibliography}

\usepackage{graphicx}	
\usepackage{amsmath}	
\usepackage{siunitx}
\usepackage{subcaption}

\title[Planet--host ratio relation]{A planet--host ratio relation to synthesize microlensing and transiting exoplanet demography from Roman}

\author[K.~Edmondson and E.~Kerins]{
Kathryn Edmondson,$^{1}$
and Eamonn Kerins$^{1}$\thanks{E-mail: Eamonn.Kerins@manchester.ac.uk}
\\
$^{1}$Department of Physics and Astronomy, University of Manchester, Oxford Road, Manchester M13 9PL, UK\\
}

\date{Accepted XXX. Received YYY; in original form ZZZ}

\pubyear{2015}

\begin{document}
\label{firstpage}
\pagerange{\pageref{firstpage}--\pageref{lastpage}}
\maketitle

\begin{abstract}
The NASA Nancy Grace Roman Space Telescope (Roman) will be the first survey able to detect large numbers of both cold and hot exoplanets across Galactic distances: $\sim$1,400 cold exoplanets via microlensing and $\sim$200,000 hot, transiting planets. Differing sensitivities to planet bulk properties between the microlensing and transit methods require relations like a planet mass--radius relation (MRR) to mediate. We propose using instead a planet--host {\em ratio} relation (PHRR) to couple directly microlensing and transit observables in demographic forward-modelling simulations. Unlike the MRR, a PHRR uses parameters that are always measured and so can potentially leverage the full Roman exoplanet sample. Using 908 confirmed exoplanets from the NASA Exoplanet Archive, we show that transit depth, $\delta$, and planet--host mass ratio, $q$, obey a PHRR that is continuous over all planet scales. The PHRR is improved by including orbital period, $P$, and host effective temperature, $T_{\star}$. We compare several candidate PHRRs of the form $\delta (q,T_\star, P)$, with the Bayesian Information Criterion favouring power-law dependence on $T_\star$ and $P$, and broken power-law dependence on $q$. The break in $q$ itself depends on $T_\star$, as do the power-law slopes in $q$ either side of the break. The favoured PHRR achieves a fairly uniform $50\%$ relative precision in $\delta$ for all $q$. Approximately $5\%$ of the sample has a transit depth that is strongly over-predicted by the PHRR; around half of these are associated with large stars ($R_\star > 2.5 \, R_{\odot}$) potentially subject to Malmquist bias.


\end{abstract}




\section{Introduction}

The exoplanet mass--radius relation is fundamental for understanding planet composition, formation and demographics, and is a key relation to consider for the design of exoplanet detection surveys. With over 5900 confirmed exoplanets in the NASA Exoplanet Archive\footnote{\url{https://exoplanetarchive.ipac.caltech.edu/}, accessed during June 2025} to date, the set of planets with both mass and radius measurements accounts for only $\sim$20\% of the dataset, motivating the need for models that enable the full dataset to be characterised. 

Given the scatter of planet mass and radius measurements, fits to the data can vary significantly depending on the selection of data, choice of model, and fit methods used. Numerous data-driven models have been presented relating exoplanet mass and radius, often employing at least two distinct power-law regimes, reflecting distinct populations of planets of differing equations of state \citep[e.g.][]{Wolfgang2016, Bashi2017, ChenKipping2017, Otegi2020, Muller2024, Parc2024}. There have also been many studies that expand the mass--radius relation to explain the large radius scatter of high mass planets, incorporating additional parameters such as insolation flux and equilibrium temperature \citep[e.g.][]{Enoch2012, Weiss2013, Thorngren2018, Sestovic2018, Ulmer-Moll2019, Edmondson2023}.

An alternative approach to traditional regression methods is that of machine learning. \citet{Tasker2020} developed a neural network based on six features and highlighted the need for predictive models that can fill in the blanks of the growing but still sparse catalogues.
\citet{Mousavi2023} performed a comprehensive analysis with machine learning, determining that orbital period was an important feature after mass and radius. They allowed clustering to classify planets rather than imposing clusters \citep[e.g. by classifying planets as rocky, volatile or gas giant as in][]{ChenKipping2017, Edmondson2023, Parc2024}.

The planet--host mass ratio, $q$, has been identified as a potentially universal parameter for planet occurrence \citep{Pascucci2018}, and is a natural parameter to use in the context of microlensing as it is retrieved directly from light-curve fitting. Additional information about the microlensing host is necessary to recover the planet mass, motivating the parameter choice of \citet{Suzuki2016}. They used 22 carefully chosen \textit{MOA-II} observations to calculate the planet--host mass ratio function for microlensing planets, finding that it changes behaviour at a characteristic mass ratio $q_{br} \sim 10^{-4}$. A similar analysis by \cite{Pascucci2018} for \textit{Kepler} exoplanets found a scaled version of the broken power law from \cite{Suzuki2016}, with $q_{br} \sim \num{3e-5}$. This relation was found to be nearly universal across host type for stars smaller than the Sun, with the exception of F-type hosts, for which the break was slightly smaller.

The Nancy Grace Roman Space Telescope (hereafter Roman) is expected to deliver a revolution in our understanding of exoplanet demographics through its Galactic Bulge Time Domain Survey \citep[GBTDS, ][]{2025arXiv250510574O}. Over a five-year nominal mission the Roman GBTDS is predicted to detect around 1,400 bound cool exoplanets \citep{Penny2019}, hundreds of free-floating planets \citep{2020AJ....160..123J} and up to 200,000 distant transiting planets \citep{2023ApJS..269....5W}. Roman will be the first survey able to detect large numbers of microlensing and transiting planets over similar Galactic distances. The differing host type and Galactic environment has been one of the primary difficulties in synthesizing demographic results from hitherto nearby transit and distant microlensing samples. By the same token, Roman will be the first survey with sensitivity that bridges both hot and cold exoplanets, enabling homogenous and fully consistent ensemble studies of the full architecture of exoplanetary systems, as well as more secure interpolation of the occurrence of planets across the entire habitable zone, a blind spot for current survey sensitivity. 

The lightcurves of microlensing and transiting events yield the planet--host mass ratio, $q$, and transit depth, $\delta$ (i.e. the square of the planet--host radius ratio), respectively. Roman is designed to secure direct host mass measurements for at least 40\% of its sample with 20\% precision or better \citep{2025arXiv251013974T}, enabling $q$ to be converted to planet mass in these cases. However, a large fraction of microlensing detections may not have well-measured host mass.

Similarly, for transiting systems, a number of brighter hosts may have well-constrained radius.  The Roman Observations Time Allocation Committee has recently endorsed the use of all available photometric and spectroscopic modes for its recommended GBTDS design in its final report \citep{2025arXiv250510574O}. Roman imaging in 8 filter bands from 0.48-2.3~$\mu$m should allow for reasonable characterisation of stellar spectral energy distributions, which will be further augmented with low-resolution grism spectroscopy that may provide better precision for brighter hosts \citep{2025arXiv250510574O}. Roman's 5-year baseline and high imaging resolution will enable very high-precision relative astrometry that can help de-blend host starlight and filter out contaminant eclipsing binary systems as they separate from target stars over the duration of the survey.

However, it is inevitable that for the many fainter hosts the host radius will be poorly determined. This means that, whilst there will be secure measurements of either $q$ or $\delta$ for all detected planets, for a large fraction of detections there will not be precise direct mass or radius measurements. 

This motivates the aim of the present paper -- to look for an empirical planet--host {\em ratio} relation (hereafter PHRR) that would provide a direct link between the microlensing and transit observables that are always measured, $q$ and $\delta$. A PHRR will enable transit and microlensing samples to be combined consistently for demographic forward models (e.g. \citet{Priyadarshi2025}).

The structure of the paper is as follows. The process for selecting a sample of confirmed planets is described in Section~\ref{sec:data}. We present and test our PHRR models against this sample in Section~\ref{sec:phrr} and summarise the results of this in Section~\ref{sec:discussion}. Concluding remarks are given in Section~\ref{sec:conclusion}.

\section{Selecting the exoplanet observational sample}\label{sec:data}

The data used for this work was obtained from the NASA Exoplanet Archive Planetary Systems Composite Data \footnote{\url{https://exoplanetarchive.ipac.caltech.edu/cgi-bin/TblView/nph-tblView?app=ExoTbls&config=PSCompPars}, accessed 8th October 2024}, using the default measurement set for each planet. At the time the data was retrieved, there were 1291 candidates with mass and radius measurements and associated uncertainties, excluding limit values. Of these, 1207 also had measurements with non-zero uncertainties in the host's mass, radius and effective temperature. Any candidate with limit values for any of these quantities was ignored. The vast majority of planet radii, with the exception of 14 direct imaging planets, are from transit observations, with transits comprising around 97\% of discoveries. Mass measurements are primarily from radial velocity follow up with a handful from imaging dynamical mass estimates or transit timing variations. None of the masses in the sample have been measured via microlensing.

Conceivably, one could imagine some statistical differences in planet bulk properties between the currently observed planet sample and a future sample from Roman based on microlensing and transit samples that will both span across much larger distances. Such differences might arise for example from differences in host metallicity distributions. These differences could alter the planet mass-radius relation, the PHRR, or indeed any other relations of planet bulk characteristics. Undoubtedly, it will be important to gain as much insight as possible on Roman host properties in order to factor in such effects. The Roman GBTDS multi-filter strategy, high astrometric sensitivity, and ability to directly detect many host lenses will play crucial roles.

In order to generate a more robust and reliable dataset, we then applied a series of selection cuts beginning with the IAU definition of an exoplanet, which imposes the conditions that $q < 2 / (25 + \sqrt{621})$ and that the planet mass should be less than the limit for deuterium burning \citep{IAU2022}, which we take to be 13~$M_J$.

We consider the plausibility of planet mass and radius pairs based on how they compare with equations of state for a planet comprised purely of iron, taken from \citet{Fortney2007}. Typically, models of terrestrial planets involve some combination of iron, rock and water, sometimes with a small atmosphere component \citep{Fortney2007, Zeng2019}. The limiting form of such a model is a planet of pure iron, and so we take this as the maximal density case. Setting the rock mass fraction to zero, the radius of such a planet, $R_{\rm Fe}$, depends on its mass $M_p$ as
\begin{eqnarray}
R_{\rm Fe} = 0.0975 (\log M_p)^2 + 0.4938 \log M_p + 0.7932,
\label{eqn:iron_radius}
\end{eqnarray}
where all quantities are in Earth units. For an equivalent expression for $M_{\rm Fe}$, we recast $M_p \rightarrow M_{\rm Fe}$ and $R_{\rm Fe} \rightarrow R_p$ in Eqn~\ref{eqn:iron_radius}, with $R_p$ the planet radius, and rearrange to get
\begin{eqnarray}
\log M_{\rm Fe} = -2.532 + 5.128 \sqrt{0.3900 R_p - 0.0655}.
\label{eqn:iron_mass}
\end{eqnarray}
Planets that are inferred to have super-iron density are excluded from our analysis.

We now transform $M_p$ and $R_p$ into $q$ and $\delta$ using
\begin{eqnarray}
q = \frac{M_p}{M_\star},
\label{eqn:mass_ratio}
\end{eqnarray}
\begin{eqnarray}
\delta = \left( \frac{R_p}{R_\star} \right)^2,
\label{eqn:transit_depth}
\end{eqnarray}
with $M_\star$ and $R_\star$ being the host mass and radius respectively. It is important to note that for the purposes of this paper, we will use ``transit depth'' and the squared ratio of planet to host radius interchangeably since we are interested in the connections between basic physical properties of planets scaled by those of their hosts. The $\delta$ we refer to here may not always equate to the observed brightness reduction in observational data due to blending or the impact of stellar limb darkening, which can serve to lower the transit floor \citep{Heller2019}. Due to the available data in the NASA Exoplanet Archive, we have calculated $\delta$ using the reported planet and host masses so that our dataset is self-consistent in this respect, but caution should be used when comparing transit depth predictions from our models to observed transit light curves without sufficient knowledge of blending effects. 

Blending occurs in crowded star fields where light from other stars contaminates the signal from the target star. In the case of transits, this acts to reduce the apparent transit depth since the blended component is not affected by the transit. This can be parameterised by a transit dilution factor, which is the fractional decrease in apparent transit depth due to blending. From pixel--level simulations of Roman's GBTDS, \citet{2023ApJS..269....5W} calculated dilution factors of around 2-4, which indicates significant deviation between true and apparent transit depths.

This is, however, the worst-case scenario as it is possible to de-blend the transit signal by observing in multiple colours over an extended time baseline. Over time, some blends will become detectably separated from the transit host due to differences in proper motion, allowing a time dependent colour change to be noticed in the host. An in-depth analysis on the projected impact of blending on Roman observations is covered by \citet{2023ApJS..269....5W}.

\begin{figure}
\includegraphics[width=0.45\textwidth]{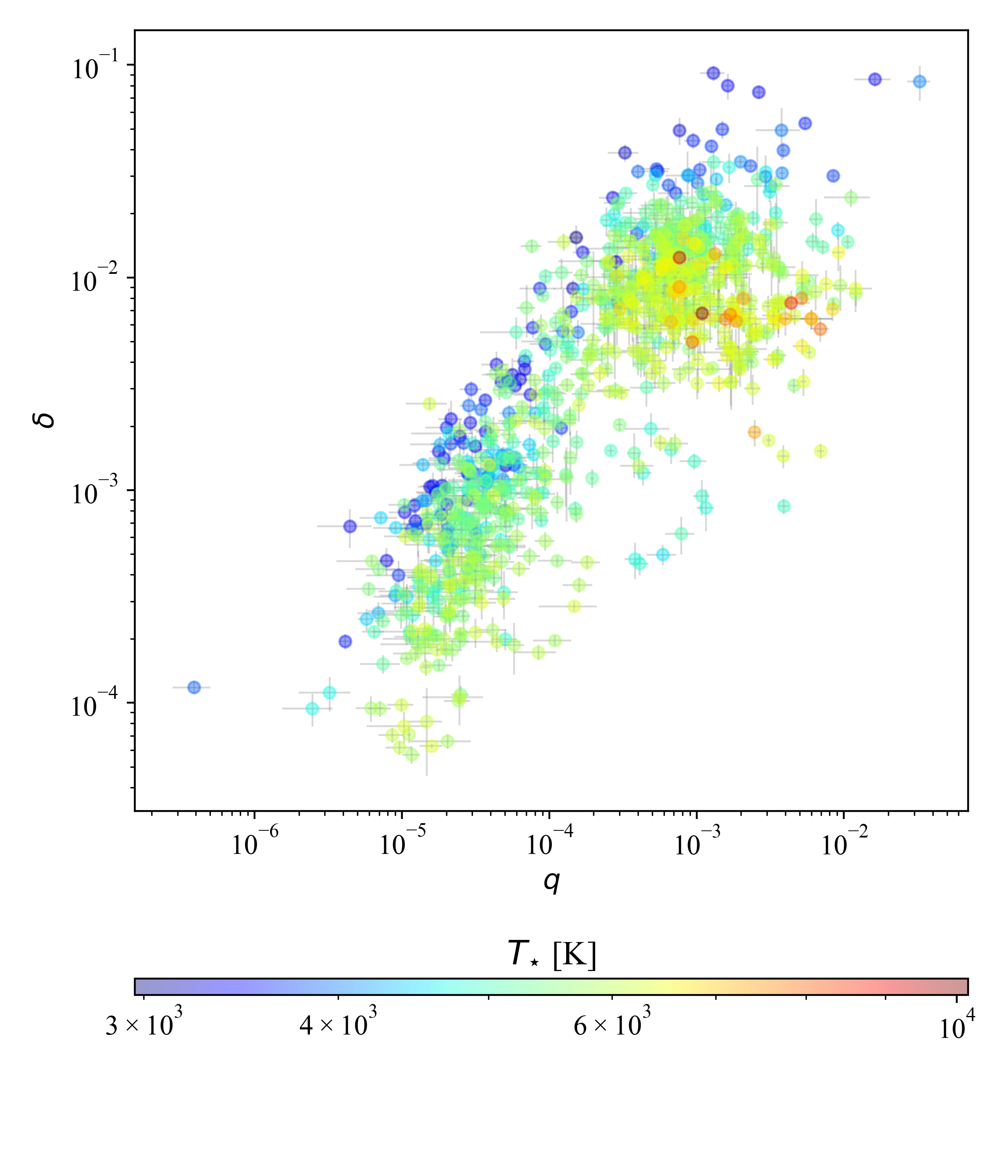}
\caption{The observed distribution of transit depth versus planet--host mass ratio for the selected sample of 986 exoplanets. The colour map depicts host effective temperature, $T_\star$, with blue to red spanning cooler to hotter hosts, respectively, on a logarithmic stretch. Stratification of the distribution with $T_\star$ is clearly evident.}
\label{fig:stteff_heatmap} 
\end{figure}

Our final selection cut imposes restrictions on data precision in log-space, such that
\begin{eqnarray}
\sigma_{\log{X}} \leq 0.2
\label{eqn:log_cut}
\end{eqnarray}
where $\sigma_{\log{X}}$ denotes the uncertainty in the base-10 logarithm of $q$, $\delta$ or host effective temperature, $T_\star$. This leaves us with a dataset of 986 exoplanets. The criteria for excluding a planet, and the number of planets excluded at each stage of the selection cuts, are summarised in Table~\ref{tab:data_cuts}. The surviving data are plotted in Figure.~\ref{fig:stteff_heatmap}.

\begin{table}
	\centering
	\caption{Summary of the exclusion criteria (inverted from the selection criteria) executed in order 1-6 with the number of candidates which failed each stage.}
	\label{tab:data_cuts}
	\begin{tabular}{lc}
		\hline
		Exclusion criteria & No. candidates removed\\
		\hline
        \\
		1. $q \geq 2 / (25 + \sqrt{621})$ & 1\\
		2. $M_p \geq 13 M_J$ & 9\\
		3. $M_p \geq M_{\rm Fe}(R_p)$ or $R_p \leq R_{\rm Fe}(M_p)$ & 23\\
        4.     $\sigma_{\log q} > 0.2$ & 145\\
        5.     $\sigma_{\log \delta} > 0.2$ & 43\\
        6.     $\sigma_{\log T_\star} > 0.2$ & 0\\
		\hline
	\end{tabular}
\end{table}

It is striking to note from Figure~\ref{fig:stteff_heatmap} how the trend of $\delta$ versus $q$ is continuous. This is in contrast to the planet mass--radius relation, which exhibits disjoint and overlapping regimes associated with rocky and ice giant systems \citep{2020A&A...634A..43O,Edmondson2023}. In $\delta$--$q$ space there is no obvious demarcation between rocky and ice-giant planets. The gradient in the $\log \delta$--$\log q$ slope is uniform from $q \sim 10^{-6}$ up to $q \sim \num{2e-4}$ before flattening out and broadening for larger $q$ where gas giant planets are increasingly dominant.

Unsurprisingly, 
Figure~\ref{fig:stteff_heatmap} also shows a clear temperature gradient over all $q$ where planets around hotter (larger) stars have smaller transit depth. This points to a PHRR having reduced scatter if host effective temperature, $T_\star$ is also incorporated. Incorporating $T_\star$ has the drawback that Roman may not be able to meaningfully measure $T_\star$ for a large fraction of hosts, potentially reducing the sample scope of the PHRR. However, incorporating $T_\star$ into the PHRR allows one to marginalise over it. In our view it is preferable to use a PHRR marginalised, where necessary, over one parameter ($T_\star$) that is statistically well studied for over a billion stars, than to connect $\delta$ and $q$ via a mass-radius relation that involves two parameters ($M_p$, $R_p$) that are jointly well measured currently for less than 1000 planets.

\begin{figure}
\centering
\includegraphics[width=0.4\textwidth]{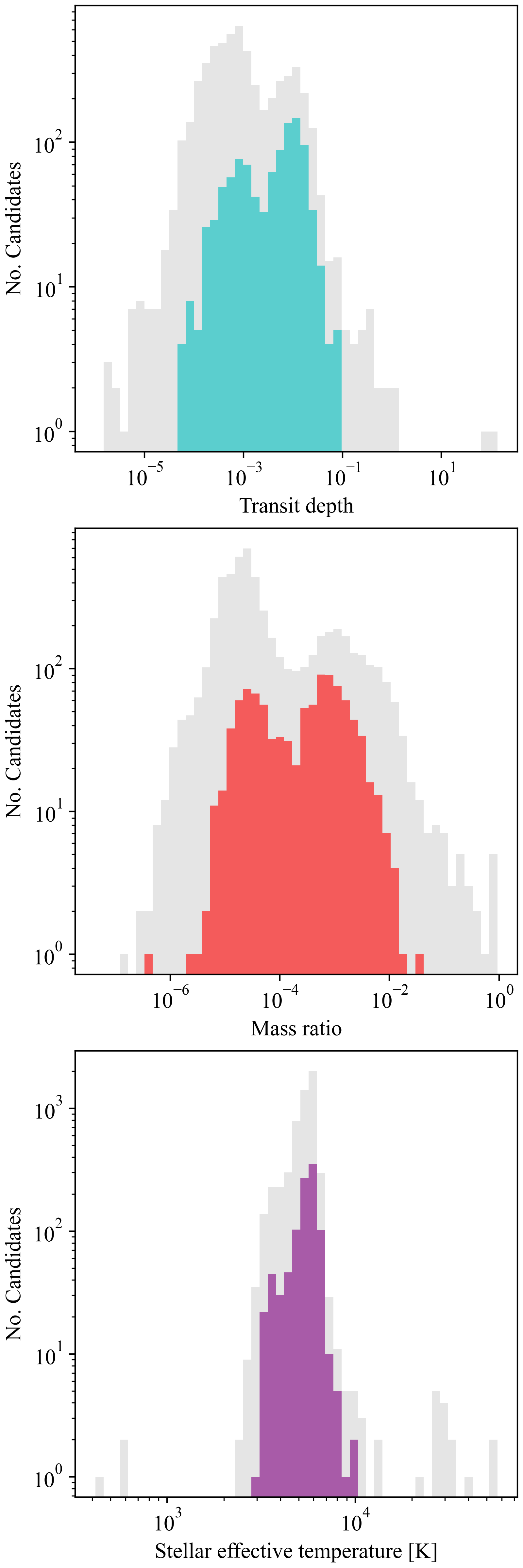}
\caption{The distributions of transit depth (top), mass ratio (middle) and stellar effective temperature (bottom) from our dataset of 986 exoplanets overlaid against the parent distributions from the NASA Exoplanet Archive.}
\label{fig:histograms} 
\end{figure}

The distributions of $q$, $\delta$ and $T_\star$ for our selected sample are compared to their respective parent distributions within the NASA Exoplanet Archive in Figure~\ref{fig:histograms}.
Both $\delta$ and $q$ distributions exhibit bimodality. Even with no completeness calculation, there is still a visible dip in occurrence at around $q \sim 10^{-4}$~-~$10^{-3}$, similar to the characteristic values observed in the planet occurrence studies of both \citet{Suzuki2016} and \citet{Pascucci2018}. The dip seen in the transit depth distribution in Figure~\ref{fig:histograms} likely corresponds to the radius valley observed by \citet{Fulton2017}.

Comparing the selected distributions to their parent distribution, it is clear that the selected sample is not an unbiased sub-sample, which is not surprising. The planet mass measurements predominately come from radial velocity observations of transiting systems. Both radial velocity and transit measurements are harder to make for smaller and lower-mass planets, making their measurement precision generally worse. Similarly, large $q$ and $\delta$ observations occur for smaller, fainter host stars that may have poorer precision in $T_\star$. Our selection rejects systems with imprecise $q$, $\delta$ or $T_\star$. Apart from at the extremes, the qualitative profiles of the selected distributions do otherwise broadly mirror their parent population.

\section{Planet--host ratio relation models}\label{sec:phrr}

\subsection{Broken power-law models}\label{sec:models}

We shall consider power-law forms for candidate PHRR. Motivated by the shape and stratification of the observed distribution in Figure~\ref{fig:stteff_heatmap}, we allow for a PHRR that scales with $T_\star$ and comprises a two-stage power-law in $q$, with a break at some $q_{br}$. We allow extra degrees of freedom in the $q>q_{br}$ regime as on visual inspection of Figure~\ref{fig:stteff_heatmap}, $\delta$ appears to fan out with $T_\star$, where as the variation for $q < q_{br}$ is well modelled by a constant shift, though we do consider a gentle quadratic variation. With these considerations in mind, the simplest form for a candidate PHRR is
\begin{eqnarray}
\delta = 10^b T_{\star}^\mu \left(\frac{q}{q_{br}}\right)^n,
\label{eqn:power_law}
\end{eqnarray}
where $b$ is a normalisation constant and $\mu$ and $n$ are model parameters that may or may not be treated as constant, depending on desired model complexity. It is more convenient to work in log space, so we transform $x \equiv \log q$, $x_{br} \equiv \log q_{br}$ and $\tau_\star \equiv \log T_\star$.

In this subsection, we consider a broken power law in two segments: subscript 1 refers to the regime $q \leq q_{br}$ and subscript 2 refers to $q > q_{br}$. To attempt to model the scatter at large $q$, we may allow the index $n_2$ to deviate from a constant and instead let it vary with stellar effective temperature as
\begin{eqnarray}
n_2 = \alpha \tau_\star + \beta,
\label{eqn:n2}
\end{eqnarray}
where $\alpha$ and $\beta$ are constants. Taking logs of Eqn~\ref{eqn:power_law}, recasting in terms of $x$ and $\tau_\star$, and substituting in Eqn~\ref{eqn:n2} gives
\begin{eqnarray}
\log \delta_2 = (\alpha \tau_\star + \beta)(x - x_{br}) + \mu \tau_\star + b.
\label{eqn:delta_2}
\end{eqnarray}
Similarly, we may allow the index $n_1$ also to vary as a different function of stellar effective temperature, with
\begin{eqnarray}
n_1 = \eta \tau_\star + \gamma,
\label{eqn:variable_n1}
\end{eqnarray}
and where $\gamma$ and $\eta$ are constants and the temperature dependence collapses when $\eta = 0$. Using superscripts $\mathrm{u}$ and $\mathrm{v}$ to represent universal ($\eta = 0$) and variable ($\eta \neq 0$) forms of $n_1$, we are left with
\begin{eqnarray}
\log \delta_{1}^{\mathrm{u}} = \gamma (x - x_{br}) + \mu \tau_\star + b
\label{eqn:delta_1u}
\end{eqnarray}
\begin{eqnarray}
\log \delta_{1}^{\mathrm{v}} = (\eta \tau_\star + \gamma) (x - x_{br}) + \mu \tau_\star + b
\label{eqn:delta_1v}
\end{eqnarray}

Finally, we consider cases where $x_{br}$ is a universal free parameter, $x_{br}^{\mathrm{u}}$, and where it is also itself a function of $T_\star$, such that
\begin{eqnarray}
x_{br}^{\mathrm{v}} = \epsilon \tau_\star + \sigma,
\label{eqn:x_br_v}
\end{eqnarray}
where the superscript $\mathrm{v}$ again stands for variable and $x_{br}^{\mathrm{u}} \equiv x_{br}^{\mathrm{v}}(\epsilon = 0)$. 

The universal and variable cases provide four combinations of models, with six to eight free parameters. All four models use Eqn~\ref{eqn:delta_2} for large $x$, then either Eqn~\ref{eqn:delta_1u} or Eqn~\ref{eqn:delta_1v} in the low $x$ regime. The separation between high and low $x$ is given by either $x_{br}$ as a constant or by Eqn~\ref{eqn:x_br_v}, where it is a function of $\tau_\star$.

We fit all four candidate PHRR models to the data using orthogonal distance regression (ODR) in \texttt{SciPy} \citep{Virtanen2020} to account for errors in both the independent variables $\log q$ and $\log T_\star$ (and later $\log P$) and the dependent variable $\log \delta$. For comparing the efficacy of PHRR models with varying numbers of parameters, we make use of the Bayesian Information Criterion (BIC), calculated from the fit $\chi^2$ value as
\begin{eqnarray}
\mathrm{BIC} = \chi^2 - k\ln N,
\label{eqn:bic}
\end{eqnarray}
where $k$ is the number of free parameters in the model and $N$ is the sample size.

\subsection{Large-star bias and outliers}\label{sec:large_star}

\begin{figure}
\includegraphics[width=0.45\textwidth]{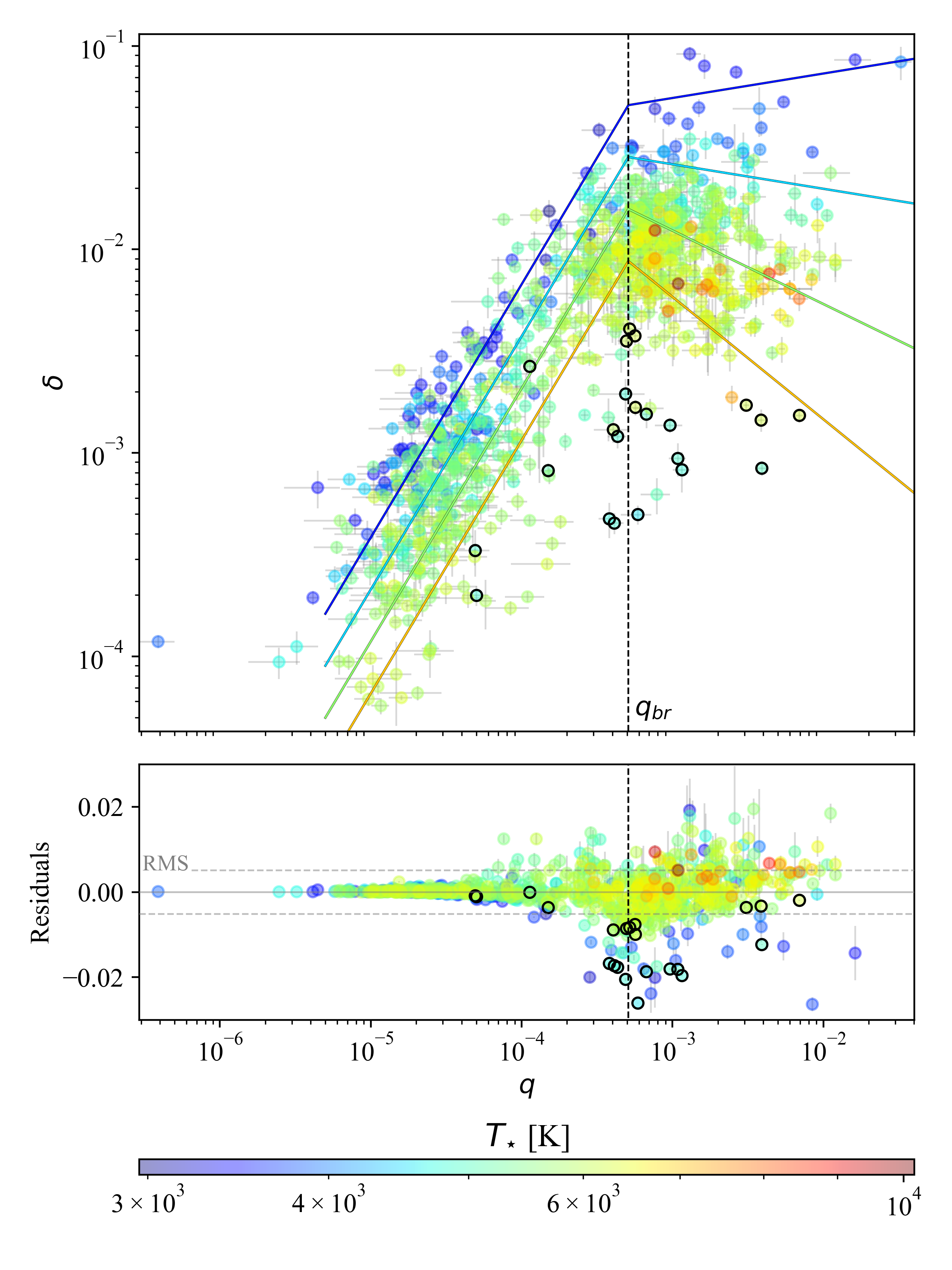}
\caption{A $q$-$\delta$ diagram for our dataset of 986 exoplanets, colour-mapped according to the effective temperature of the host star. The ODR best-fit model with a universal $n_1$ and universal $q_{br}$ is plotted for several representative values of $T_\star$, where the colour of the line also corresponds to the legend at the bottom. The bottom panel shows the residuals (measured minus predicted where the prediction for each planet is calculated using its measured value of $T_\star$) along with the RMS residual (grey dashed line). 22 planets that orbit large stars ($R_\star \gtrsim 2.5 R_\odot$) are marked in both panels by black rings.}
\label{fig:large_star_q_delta} 
\end{figure}

Figure~\ref{fig:large_star_q_delta} shows the result of fitting a six-parameter PHRR model with universal forms for $n_1$ and $q_{br}$. The model provides a reasonable level of agreement for the large majority of the sample, though it is evident that it systematically over-predicts transit depth for a small fraction ($\sim 5\%$) of planets with $q$ in the neigbourhood of $q_{br}$. Similar behaviour is evident for all of the PHRR models.

\begin{figure}
\includegraphics[width=0.45\textwidth]{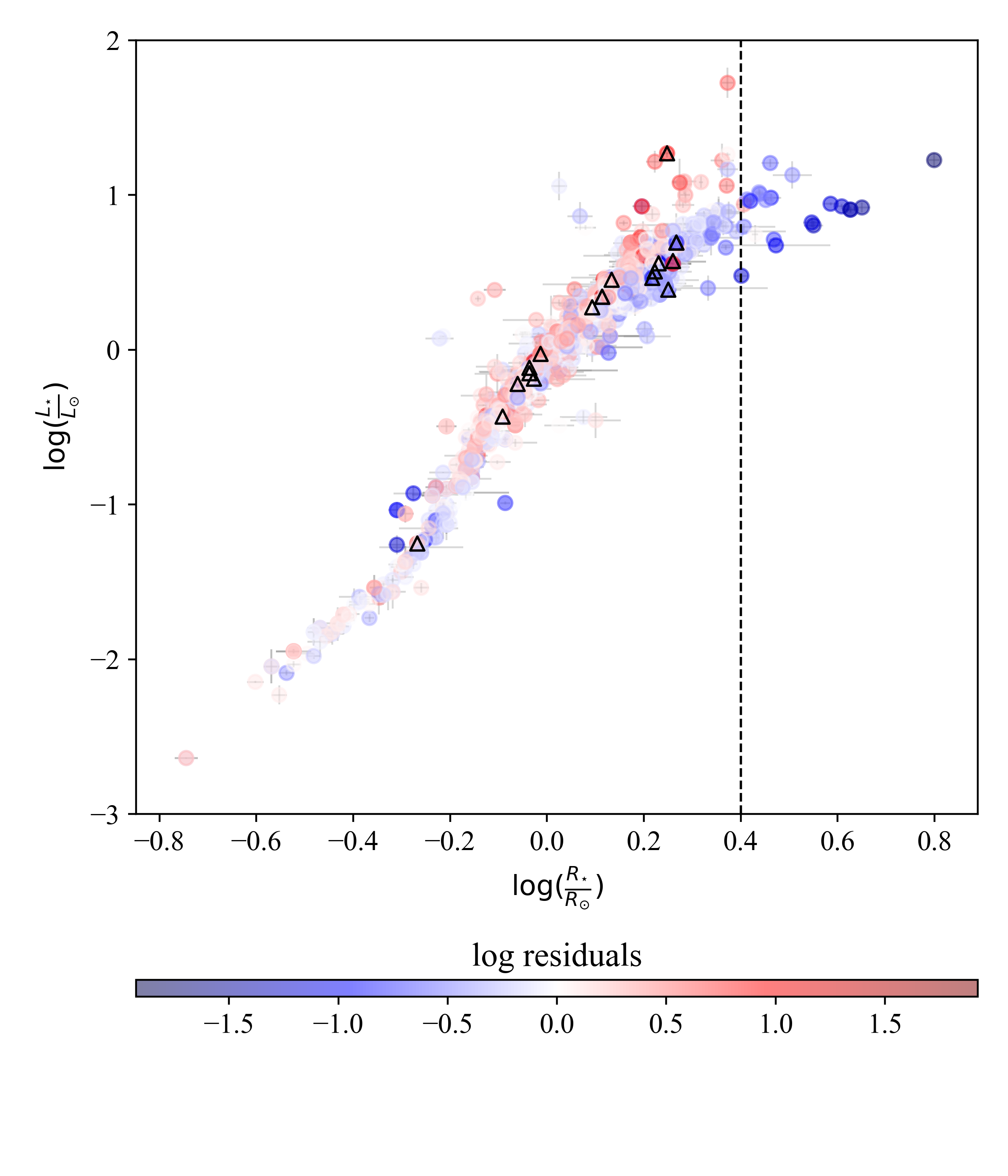}
\caption{Host luminosity plotted against host radius for each of the 986 planets in our dataset.The colour scale represents the log residuals (log transit depth subtract model prediction in log-space) from the universal $n_1$ and universal $q_{br}$ model in Figure~\ref{fig:large_star_q_delta} such that blue points are over-predictions and red points are under-predictions. Our chosen large-star threshold at $\log \left( \frac{R_\star}{R_\odot} \right) = 0.4$ is shown by the black dashed line. Hosts which have subsequently been removed by iterative sigma clipping and fall below the large star threshold are denoted by open black triangles -- this has removed more than one planet for some hosts hence there are less than 22 outliers visibly plotted here.}
\label{fig:large_star_rstar_lum} 
\end{figure}

To investigate a potential cause of this behaviour the luminosity and stellar radii of the host stars have been plotted in Figure~\ref{fig:large_star_rstar_lum} with a heat map to highlight the residuals from the bottom panel of Figure~\ref{fig:large_star_q_delta}. It is clear from Figure~\ref{fig:large_star_q_delta} that a number of the largest outliers are associated with large host stars with $R_\star > 2.5\, R_{\odot}$.

Such a discrepancy is consistent with Malmquist bias, the observational tendency to detect intrinsically bright objects as a consequence of magnitude limited surveys. \citet{Gaidos2013} demonstrate that this is especially evident in transit surveys, as while the sampling volume increases as $d^3$ for a survey that includes objects out to distance $d$, there is an additional factor of $d$ introduced by the geometric probability of transit. Given large uncertainties in stellar radii, they found that many planets around larger stars may be larger than they appear to be by up to a factor of two. This is consistent with the behaviour seen in Figure~\ref{fig:large_star_rstar_lum} - the PHRR model over-predicts for many large stars as it expects a deeper transit than was observed. While much of the Kepler catalogue has been corrected to account for this through the use of Gaia parallaxes \citep{Berger2018}, it seems plausible that the problem may still persist for a small minority of cases. We therefore choose to remove planets that orbit host stars larger than $10^{0.4} R_\odot \sim 2.5 R_\odot$ based upon visual inspection of Figure~\ref{fig:large_star_rstar_lum}, as shown by the dashed line. The 22 planets affected by this cut are highlighted by black rings in Figure~\ref{fig:large_star_q_delta} and were removed from the dataset for any subsequent analysis, leaving a sample of 964 planets.

\begin{figure*}
\includegraphics[width=1.0\textwidth]{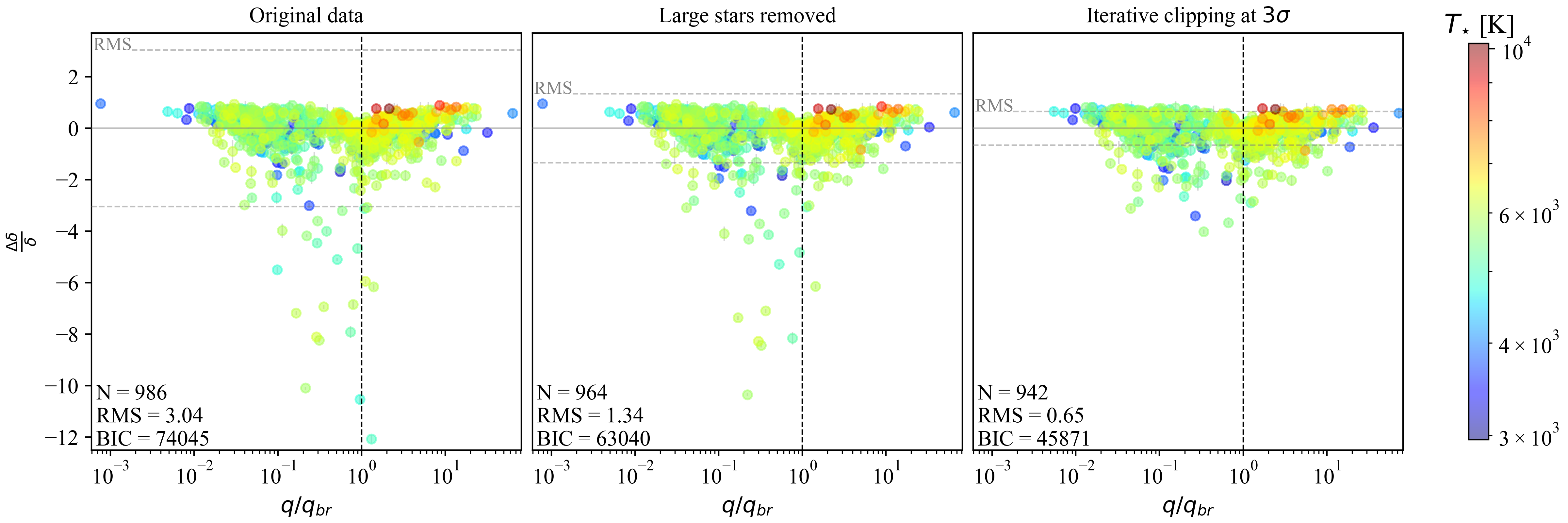}
\caption{Relative residuals in transit depth (observed minus predicted over observed) plotted against the normalised mass-ratio ($q / q_{br}$) for the PHRR model with universal $n_1$ and universal $q_{br}$. Error bars are plotted but are often smaller than the size of the data point symbols. Data is colour-mapped according to $T_\star$. The left panel uses the original dataset of $N = 986$ exoplanets selected in Section~\ref{sec:data}, the middle panel shows the effect of removing large stars before fitting, and the right panel demonstrates the impact of iterative sigma clipping at $3\sigma$ following the removal of large stars. The RMS relative residual is plotted (grey dashed lines) and displayed on each panel along with the Bayesian Information Criterion.}
\label{fig:outliers_comparison} 
\end{figure*}

To handle any remaining outliers, we made use of iterative sigma clipping, whereby data points were considered as outliers and removed on a given iteration if they were more than $3\sigma$ away from the median residuals, where $\sigma$ is the standard deviation of the residuals. The PHRR model was then re-fit, and the process repeated until the dataset and model yielded no outliers. The impact both of removing large stars and of iterative sigma clipping is demonstrated in Figure~\ref{fig:outliers_comparison}.

The RMS relative residual improved by more than a factor of two upon the removal of large stars and the BIC improved by over 11000, though the over-prediction tail around $q = q_{br}$ is still present in the middle panel of Figure~\ref{fig:outliers_comparison}. Iterative clipping at $3 \sigma$ removes a further 22 planets and again halves the RMS relative residual and significantly improves the BIC score. Still, it is evident from the residuals that there is a tendency to over-predict transit depth for a small minority of planets located near, but largely below, $q_{br}$. This could indicate a systematic deviation of the PHRR from pure power-law behaviour in the regime $q \leq q_{br}$, which we now investigate.

\subsection{Moving beyond a power law}\label{sec:quadratic}

Figure~\ref{fig:outliers_comparison} demonstrates relatively uniform scatter of around 50\% above $q_{br}$, which is competitive with the scatter seen in the mass-radius relation (c.f. Figure~\ref{fig:phpp_norm_resids}). We therefore do not expect any significant further improvements from a more complex PHRR. We will, however, test whether or not additional degrees of freedom in the $q\leq q_{br}$ regime can better describe the potentially systematic over-prediction tail towards $q_{br}$. To attempt to better model the $q \leq q_{br}$ regime we extend the PHRR with a quadratic term in $q$, and with a quadratic coefficient that is allowed to vary with $T_\star$. Using superscript $\mathrm{Q}$ to represent this quadratic model and subscript 1 as before to indicate applicability only to $q \leq q_{br}$, we modify Eqn~\ref{eqn:delta_1u} to
\begin{eqnarray}
\log \delta_{1}^{\mathrm{Q}} = \gamma (x - x_{br}) + \mu \tau_\star + b + (\lambda - \tau_\star) ( x - x_{br})^2,
\label{eqn:delta_1q}
\end{eqnarray}
where $\lambda$ is a constant. As the over-predictive behaviour seems more prominent at lower values of $T_\star$, we scale the quadratic term negatively with $\tau_\star$.

\begin{figure}
\includegraphics[width=0.45\textwidth]{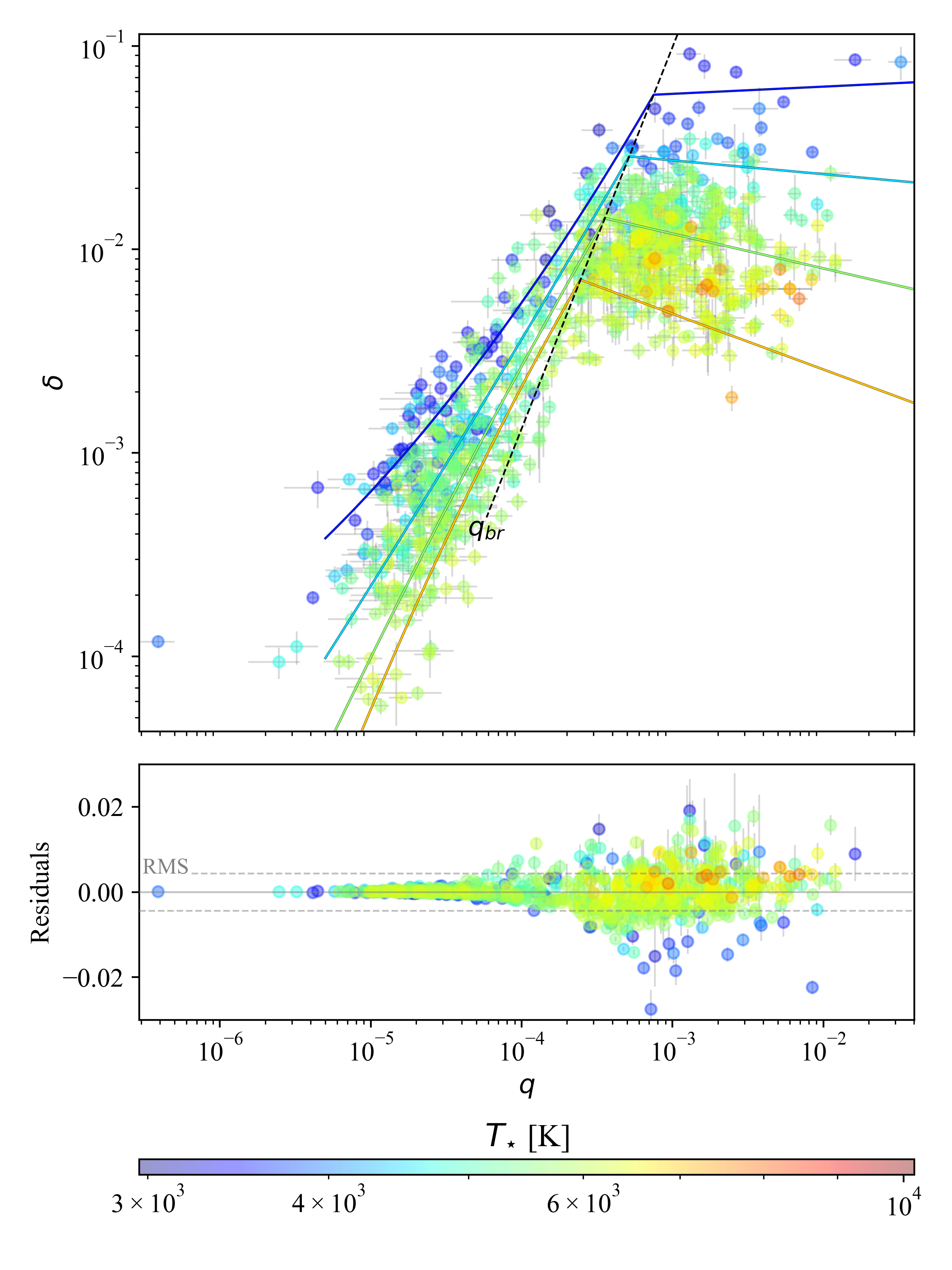}
\caption{The $q$-$\delta$-$T_\star$ distribution, as shown in Figure~\ref{fig:large_star_q_delta}, fitted to the quadratic PHRR model with variable ($T_\star$ dependent) $q_{br}$.}
\label{fig:qv_q_delta} 
\end{figure}

A list of the fit parameters and variables used in all of our models is laid out in Table~\ref{tab:param_defs}, including those relevant to the treatment of orbital period discussed later in Section~\ref{sec:orb_period}. Table~\ref{tab:param_results} summarises the fit parameters, BIC, and number of planets removed due to iterative clipping at $3 \sigma$, for the six PHRR models presented so far. Large host stars and outliers identified by sigma clipping at $3\sigma$ were removed. The quadratic model with temperature-dependent $q_{br}$ is shown in Figure~\ref{fig:qv_q_delta} as an example.

\begin{table}
    \centering
    \caption{Summary of the key fit variables used for the suite of PHRR models considered in this paper. These variables are provided in equations throughout Section~\ref{sec:phrr}.}
    \label{tab:param_defs}
    \begin{tabular}{ll}
        \hline
        \multicolumn{2}{l}{\textbf{Variables}}\\
        \hline
        $\log \delta$ & Base-10 logarithm of transit depth from Eqn~\ref{eqn:transit_depth}\\
        $x$ & Base-10 logarithm of mass ratio from Eqn~\ref{eqn:mass_ratio}\\
        $\tau_{\star}$ & Base-10 logarithm of the host effective temperature in K\\
        $p$ & Base-10 logarithm of the orbital period in days\\
        \hline
        \multicolumn{2}{l}{\textbf{Fit Parameters}}\\
        \hline
         $\gamma$ &  Offset of the mass-ratio index for $q\leq q_{br}$\\
         $x_{br}$ &  Base-10 logarithm of the characteristic mass ratio\\
         $\mu$ & Host effective temperature power-law index\\
         $b$ &  Normalisation constant -- an offset in log space\\
         $\alpha$ & Strength of $T_\star$ dependence for the mass ratio slope for $q>q_{br}$\\
         $\beta$ &  Offset of the mass-ratio index for $q> q_{br}$\\
         $\epsilon$ &  Strength of $T_\star$ dependence for $q_{br}$\\
         $\sigma$ &  Offset for $q_{br}$ in log space\\
         $\eta$ &  Strength of $n_1$ dependence on $T_\star$\\
         $\lambda$ & Strength of quadratic behaviour for $q\leq q_{br}$\\
         $\kappa$ &  Orbital period power law index\\
         \hline
    \end{tabular}
\end{table}

\begin{table*}
	\centering
	\caption{Best-fit parameters and uncertainties, BIC values and the number of data points removed during iterative sigma clipping at $3\sigma$, $N_{clip}$, for all $q$-$\delta$-$T_\star$ models.}
	\label{tab:param_results}
	\begin{tabular}{ccccccc}
		\hline
            & \multicolumn{2}{c}{Universal $n_1$} & \multicolumn{2}{c}{Variable $n_1$} & \multicolumn{2}{c}{Quadratic}\\
            & Universal $q_{br}$ & Variable $q_{br}$ & Universal $q_{br}$ & Variable $q_{br}$ & Universal $q_{br}$ & Variable $q_{br}$\\
        \hline
            $N_{clip}$ & 22 & 21 & 26 & 25 & 26 & 26\\
            BIC & 45871 & 46444 & 43690 & 44718 & 44384 & 46069\\
		\hline
            $\gamma$ & $1.270 \pm 0.008$ & $1.265 \pm 0.008$ & $0.9 \pm 0.3$ & $1.750 \pm 0.09$ & $1.50 \pm 0.03$ & $1.28 \pm 0.03$\\
            $x_{br}$ & $-3.347 \pm 0.006$ & - & $-3.358 \pm 0.006$ & - & $-3.512 \pm 0.007$ & -\\
            $\mu$ & $-2.59 \pm 0.04$ & $-2.58 \pm 0.06$ & $-2.51 \pm 0.06$ & $-2.56 \pm 0.06$ & $-1.90 \pm 0.04$ & $-3.03 \pm 0.06$\\
            $b$ & $7.9 \pm 0.2$ & $7.9 \pm 0.2$ & $7.6 \pm 0.2$ & $7.8 \pm 0.2$ & $5.2 \pm 0.2$ & $9.5 \pm 0.2$\\
            $\alpha$ & $-1.69 \pm 0.08$ & $-1.8 \pm 0.1$ & $-1.8 \pm 0.1$ & $-1.7 \pm 0.1$ & $-2.30 \pm 0.08$ & $-1.0 \pm 0.1$\\
            $\beta$ & $6.1 \pm 0.3$ & $6.4 \pm 0.4$ & $6.5 \pm 0.4$ & $6.3 \pm 0.4$ & $8.5 \pm 0.3$ & $3.7 \pm 0.4$\\
            $\epsilon$ & - & $-0.05 \pm 0.06$ & - & $-0.44 \pm 0.09$ & - & $-1.62 \pm 0.08$\\
            $\sigma$ & - & $-3.2 \pm 0.2$ & - & $-1.7 \pm 0.3$ & - & $2.6 \pm 0.3$\\
            $\eta$ & $\equiv0$ & $\equiv0$ & $0.09 \pm 0.07$ & $0.5 \pm 0.1$ & - & -\\
            $\lambda$ & - & - & - & - & $3.80 \pm 0.02$ & $3.68 \pm 0.01$\\
		\hline

	\end{tabular}
\end{table*}

The BIC values favour a universal mass ratio break when comparing quadratic models, however neither of these can improve upon the model with variable $n_1$ and universal $q_{br}$. The residuals in Figure~\ref{fig:qv_q_delta} are scattered at large mass ratios for both under- and over-predictions. Over-predicted transit depths beyond the RMS residual are almost exclusively for the coolest stars, which could be indicative of a failure to fully characterise the transition regime between low and high mass ratios.

It is worth noting that models with universal $n_1$ required fewer planets to be removed as outliers due to iterative sigma clipping, though the discrepancy is small compared to the BIC differences, which are of order $\sim$1000 in favour of the variable $n_1$ models. The BIC favours a variable $n_1$ and universal $q_{br}$, which further evidences a lack of improvement by introducing a quadratic term. For models where $x_{br}$ is treated as universal BIC scores are similar, which supports the idea that this feature is characteristic of the PHRR. For variable $n_1$ models, the value of $\eta$ is modestly negative, indicating that scaling with $\tau_\star$ is relatively weak. Nonetheless, the BIC scores are smaller for these models, implying some small improvement.

The temperature index $\mu$ is also fairly consistent across all models at around -2.6, demonstrating a strong inverse relationship between transit depth and stellar effective temperature, all else being equal. The Stefan-Boltzmann law ($L_\star \propto R_\star^2 T_\star^4$) coupled with a main sequence stellar mass--luminosity relation of approximately $L_\star \propto M_\star^4$ suggests a scaling of $T_\star \propto \delta^{0.25} q^{-1}$ for fixed planet bulk properties. On the other hand, a PHRR of the form of Eq~(\ref{eqn:power_law}) predicts $T_\star \propto \delta^{1/\mu} q^{-n/\mu}$. From Table~\ref{tab:param_results} and Eq~(\ref{eqn:variable_n1}) we see that, e.g. for the PHRR with universal $n_1$ and $q_{br}$, we have $\mu = -2.59$ and $n_1 = \gamma = 1.27$, giving $T_\star \propto \delta^{-0.39} q^{0.49}$. The strong difference between this prediction and the expectation from consideration of the stellar mass-luminosity relation is a measure of the differences in typical observed planet bulk characteristics as $T_\star$ increases. These differences may be due to detection bias, or to a combination of detection bias and intrinsic variation in planet bulk properties with different hosts.

The extent of quadratic behaviour is governed by $\lambda$ in Equation~\ref{eqn:delta_1q} such that if $\lambda \sim \tau_\star$, the model will collapse back to the universal $n_1$ form. The mean and standard deviation of $\tau_\star$ is $3.73 \pm 0.07$, which is consistent with both values of $\lambda$ in Table~\ref{tab:param_results}. Considering the fit uncertainty on $\lambda$, 49\% of $\tau_\star$ values fall within $3 \sigma$ of $\lambda$ for the universal $q_{br}$ model and 12\% for the variable $q_{br}$ model. The quadratic component of the model therefore plays a small role in describing the dependence of transit depth on mass ratio and stellar effective temperature, as is evident in Figure~\ref{fig:qv_q_delta}, where best fit lines only appear to demonstrate any noticeable curvature for the coolest stars. This may just be an artifact of relatively small number statistics for cool stars, or the increasing effect of outlier behaviour near $q_{br}$ for all models, rather than any true non-linear behaviour at low $T_\star$. We will explore how we might reduce the residual scatter and tackle the systematic over-predictions of transit depth for the coolest stars in the next section.

\subsection{Accounting for orbital period}\label{sec:orb_period}

We now investigate if the remaining scatter for planets with cooler host stars results from marginalising over orbital period, or if it is more likely to be intrinsic. We carry forward the data selection process in Section~\ref{sec:data} with some additional constraints:
\begin{enumerate}
    \item we require non-zero errors and central measurements in orbital period now as well, which reduces the dataset by 55 candidates.
    \item we require precision in $\log P$ as in Eqn~\ref{eqn:log_cut}, which removes a further 1 candidate.
\end{enumerate}

\begin{figure}
\centering
\includegraphics[width=0.4\textwidth]{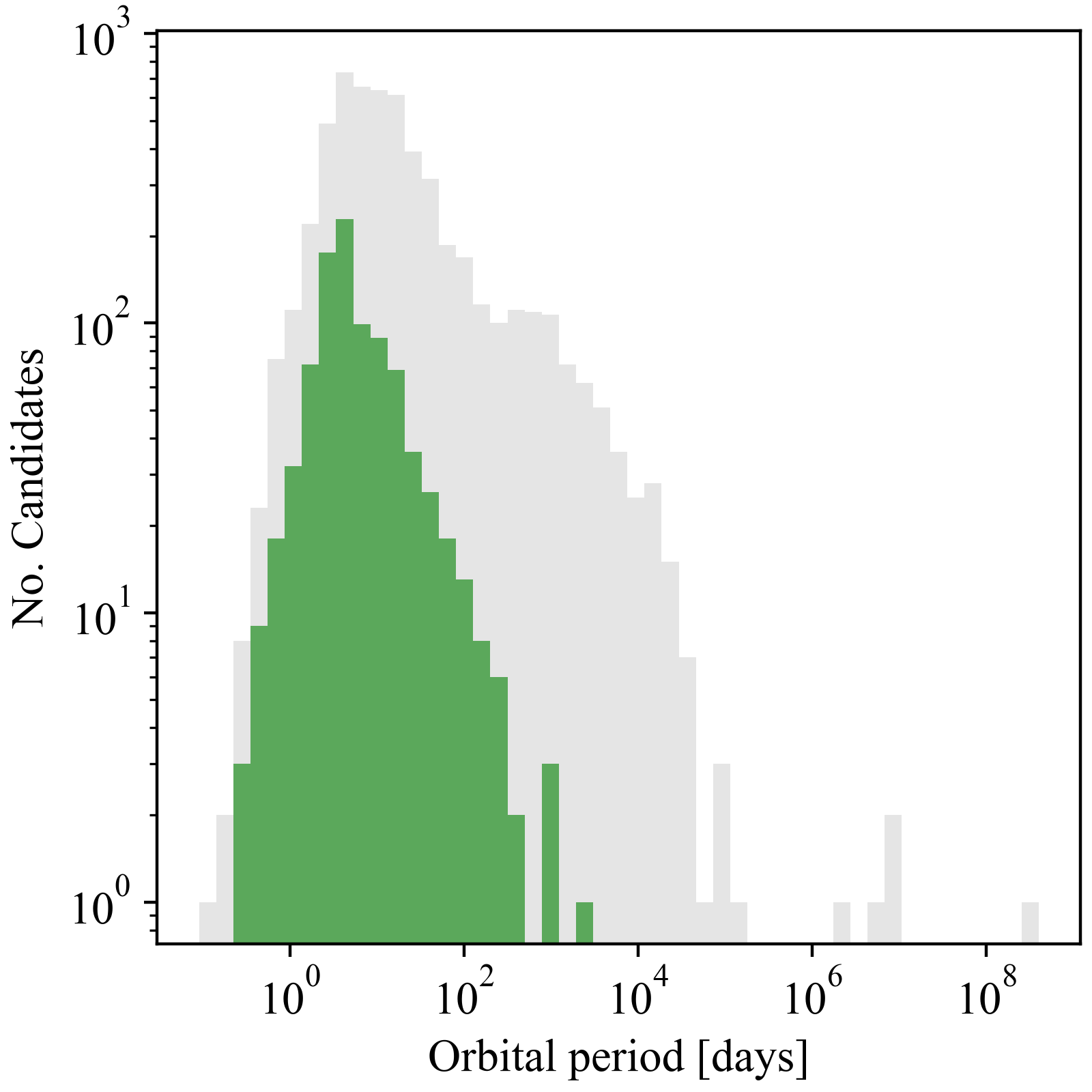}
\caption{The distribution of orbital period for our dataset now consisting of 908 exoplanets overlaid against the parent distribution from the NASA Exoplanet Archive.}
\label{fig:histograms_period} 
\end{figure}

The period distribution for our revised dataset is plotted in Figure~\ref{fig:histograms_period}.
For short periods, our distribution is very close to the parent distribution, indicating that the majority of short-period planets pass our selection cuts. As orbital period increases, the discrepancy between the distributions widens, which can be reconciled by the requirement for measurements in both mass ratio and transit depth. The vast majority of our data consists of transiting exoplanets with radial velocity follow up - both of these methods are more sensitive to short period planets. 

We attempt to capture any dependence on orbital period in a single power law, for which we add and extra term to each model both above and below $q_{br}$ - that is, we add $\kappa p$ to Eqns \ref{eqn:delta_2}, \ref{eqn:delta_1u}, \ref{eqn:delta_1v} and \ref{eqn:delta_1q} where $p \equiv \log P$ and $\kappa$ is an additional free parameter of the fit. For the non-quadratic PHRR models, this means that transit depth now scales as
\begin{eqnarray}
\delta \propto T_\star^\mu P^\kappa \left( \frac{q}{q_{br}}\right)^n 
\label{eqn:power_law_p}
\end{eqnarray}
where again $n$ and $q_{br}$ are not necessarily constants but can depend on $T_\star$.

As in the previous sections, large host stars and outliers were excluded in the fit. The six models developed in previous sections are adjusted to include this power-law orbital period dependence and again fit to the data using ODR. The results of the fit, BIC and number of outliers identified are presented in Table~\ref{tab:param_results_period}.

\begin{table*}
	\centering
	\caption{Similar to Table~\ref{tab:param_results} for all $q$-$\delta$-$T_\star$-$P$ models. The fit parameters for the BIC favoured model are shown in bold.}
	\label{tab:param_results_period}
	\begin{tabular}{ccccccc}
		\hline
		  & \multicolumn{2}{c}{Universal $n_1$} & \multicolumn{2}{c}{Variable $n_1$} & \multicolumn{2}{c}{Quadratic}\\
            & Universal $q_{br}$ & Variable $q_{br}$ & Universal $q_{br}$ & Variable $q_{br}$ & Universal $q_{br}$ & Variable $q_{br}$\\
        \hline
            $N_{clip}$ & 24 & 24 & 24 & 26 & 26 & 27\\
            BIC & 39601 & 39386 & 39690 & 38406 & 40038 & 40215\\
		\hline
            $\gamma$ & $1.311 \pm 0.009$ & $1.307 \pm 0.009$ & $1.5 \pm 0.3$ & $\mathbf{-2.3 \pm 0.4}$ & $1.45 \pm 0.03$ & $1.03 \pm 0.04$\\
            $x_{br}$ & $-3.388 \pm 0.006$ & - & $-3.395 \pm 0.006$ & - & $-3.538 \pm 0.008$ & -\\
            $\mu$ & $-2.69 \pm 0.05$ & $-2.75 \pm 0.07$ & $-2.69 \pm 0.7$ & $\mathbf{-3.66 \pm 0.08}$ & $-1.91 \pm 0.04$ & $-3.12 \pm 0.07$\\
            $b$ & $8.3 \pm 0.2$ & $8.5 \pm 0.2$ & $8.3 \pm 0.2$ & $\mathbf{11.9 \pm 0.3}$ & $5.3 \pm 0.2$ & $9.9 \pm 0.3$\\
            $\alpha$ & $-1.12 \pm 0.08$ & $-1.0 \pm 0.1$ & $-1.1 \pm 0.1$ & $\mathbf{0.5 \pm 0.1}$ & $-1.86 \pm 0.08$ & $-0.6 \pm 0.1$\\
            $\beta$ & $4.0 \pm 0.3$ & $3.6 \pm 0.4$ & $4.0 \pm 0.3$ & $\mathbf{-2.0 \pm 0.5}$ & $6.9 \pm 0.3$ & $2.1 \pm 0.5$\\
            $\epsilon$ & - & $-0.19 \pm 0.06$ & - & $\mathbf{-1.9 \pm 0.1}$ & - & $-1.61 \pm 0.08$\\
            $\sigma$ & - & $-2.67 \pm 0.06$ & - & $\mathbf{3.5 \pm 0.4}$ & - & $2.6 \pm 0.3$\\
            $\eta$ & $\equiv0$ & $\equiv0$ & $-0.06 \pm 0.08$ & $\mathbf{1.0 \pm 0.1}$ & - & -\\
            $\lambda$ & - & - & - & - & $3.74 \pm 0.02$ & $3.50 \pm 0.02$\\
            $\kappa$ & $-0.094 \pm 0.004$ & $-0.096 \pm 0.004$ & $-0.095 \pm 0.004$ & $\mathbf{-0.102 \pm 0.004}$ & $-0.111 \pm 0.004$ & $-0.109 \pm 0.004$\\
		\hline

	\end{tabular}
\end{table*}

Across the board, the BIC in Table~\ref{tab:param_results_period} are significantly smaller than those in Table~\ref{tab:param_results}, which clearly demonstrates improvement in describing transit depth when power law dependence on orbital period is included. There is also good agreement across models that the power law index describing the period dependence is around $-0.10$, which is weak but non-zero. The majority of parameters for the $q$-$\delta$-$T_\star$ part of the model are consistent with the same behaviour as before, with slightly higher uncertainties due to the introduction of an additional free parameter. An exception to this is the high mass ratio behaviour of the variable--$n_1$--variable--$q_{br}$ model, which is the only model with $\alpha > 0$. This would indicate an increase in the mass ratio index with temperature. The value of $\epsilon$ is more significantly negative for this model than without period dependence, providing stronger evidence in favour of an inverse relationship between characteristic mass ratio and stellar effective temperature.

\begin{figure}
\includegraphics[width=0.45\textwidth]{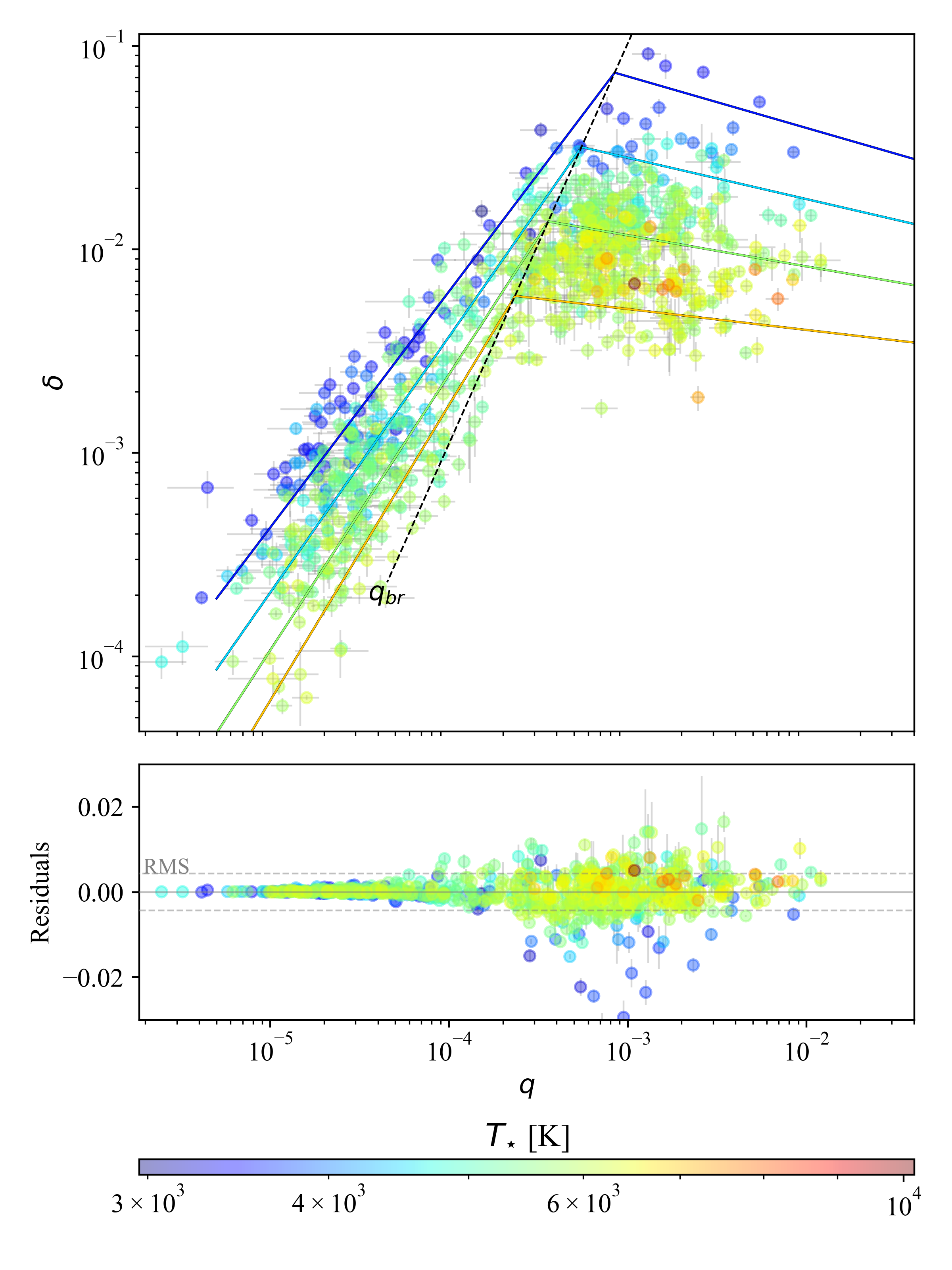}
\caption{PHRR as in Figures \ref{fig:large_star_q_delta} and \ref{fig:qv_q_delta} for the model with variable $n_1$, variable $q_{br}$, and power-law orbital period dependence. Based on BIC score, this is the most favoured model out of all considered in this paper. The model lines at various $T_\star$ values are plotted for the logarithmically averaged period of 5.6~days.}
\label{fig:vvp_q_delta} 
\end{figure}

The BIC-favoured model is plotted in Figure~\ref{fig:vvp_q_delta} and allows both $n_1$ and $q_{br}$ to vary with $T_\star$. The favoured model takes the form
\begin{eqnarray}
\delta = 0.012 \left( \frac{q}{q_{br}} \right)^n \left(\frac{T_\star}{T_\odot} \right)^{-3.7} \left(\frac{P}{10\, \mathrm{days}} \right)^{-0.10} \\
n = \begin{cases}
1.0\log{\left(\frac{T_\star}{T_\odot}\right)}+1.4 & q \leq q_{br} \\
0.51\log{\left(\frac{T_\star}{T_\odot}\right)}-0.15 & q > q_{br}
\end{cases}\\
q_{br} = \num{3.4e-4}\left(\frac{T_\star}{T_\odot} \right)^{-1.9},
    \label{eqn:phrr_with_params_scaled}
\end{eqnarray}
where the bold parameters from Table~\ref{tab:param_results_period} have been substituted into Equations~\ref{eqn:n2}, \ref{eqn:variable_n1}, \ref{eqn:x_br_v} and \ref{eqn:power_law_p}. The full covariance matrix for this model is presented in Table~\ref{tab:cov_matrix}.

\begin{table*}
    \centering
    \caption{The covariance matrix for the fit parameters of the favoured PHRR model (variable--$n_1$--variable--$q_{br}$). The diagonal elements are in bold. Note the large magnitude of many off-diagonal elements compared to the diagonal ones indicative of significant covariance.}
    \label{tab:cov_matrix}
    \begin{tabular}{rccccccccc}
        \hline
         & $\gamma$ & $\epsilon$ & $\sigma$ & $\mu$ & $b$ & $\alpha$ & $\beta$ & $\eta$ & $\kappa$\\
        \hline
        $\gamma$ & \textbf{\num{1.4e-1}} & & & & & & & & \\
        $\epsilon$ & \num{2.6e-2} & \textbf{\num{1.0e-2}} & & & & & & & \\
        $\sigma$ & \num{-9.5e-2} & \num{-3.9e-2} & \textbf{\num{-1.5e-1}} & & & & & & \\
        $\mu$ & \num{-5.7e-3} & \num{3.1e-3} & \num{-1.1e-2} & \textbf{\num{5.7e-3}} & & & & & \\
        $b$ & \num{2.1e-2} & \num{-1.2e-2} & \num{4.4e-2} & \num{-2.1e-2} & \textbf{\num{8.0e-2}} & & & & \\
        $\alpha$ & \num{1.9e-3} & \num{-7.0e-3} & \num{2.6e-2} & \num{-8.8e-3} & \num{3.3e-2} & \textbf{\num{1.8e-2}} & & & \\
        $\beta$ & \num{-7.3e-3} & \num{2.6e-2} & \num{-9.9e-2} & \num{3.3e-2} & \num{-1.2e-1} & \num{-6.8e-2} & \textbf{\num{2.6e-1}} & & \\
        $\eta$ & \num{-3.7e-2} & \num{-6.7e-3} & \num{2.6e-2} & \num{1.5e-3} & \num{-5.8e-3} & \num{-5.2e-4} & \num{2.0e-3} & \textbf{\num{9.9e-3}} & \\
        $\kappa$ & \num{1.7e-5} & \num{-3.5e-6} & \num{1.8e-5} & \num{-2.0e-5} & \num{-6.7e-5} & \num{1.5e-6} & \num{3.1e-7} & \num{-5.6e-6} & \textbf{\num{1.5e-5}}\\
        \hline
    \end{tabular}
\end{table*}

There is strong covariance between several parameters in Table~\ref{tab:cov_matrix}, therefore if propagating uncertainties in $\delta$ via Equation~\ref{eqn:phrr_with_params_scaled}, it is necessary to account for this.

\begin{figure}
    \includegraphics[width=0.45\textwidth]{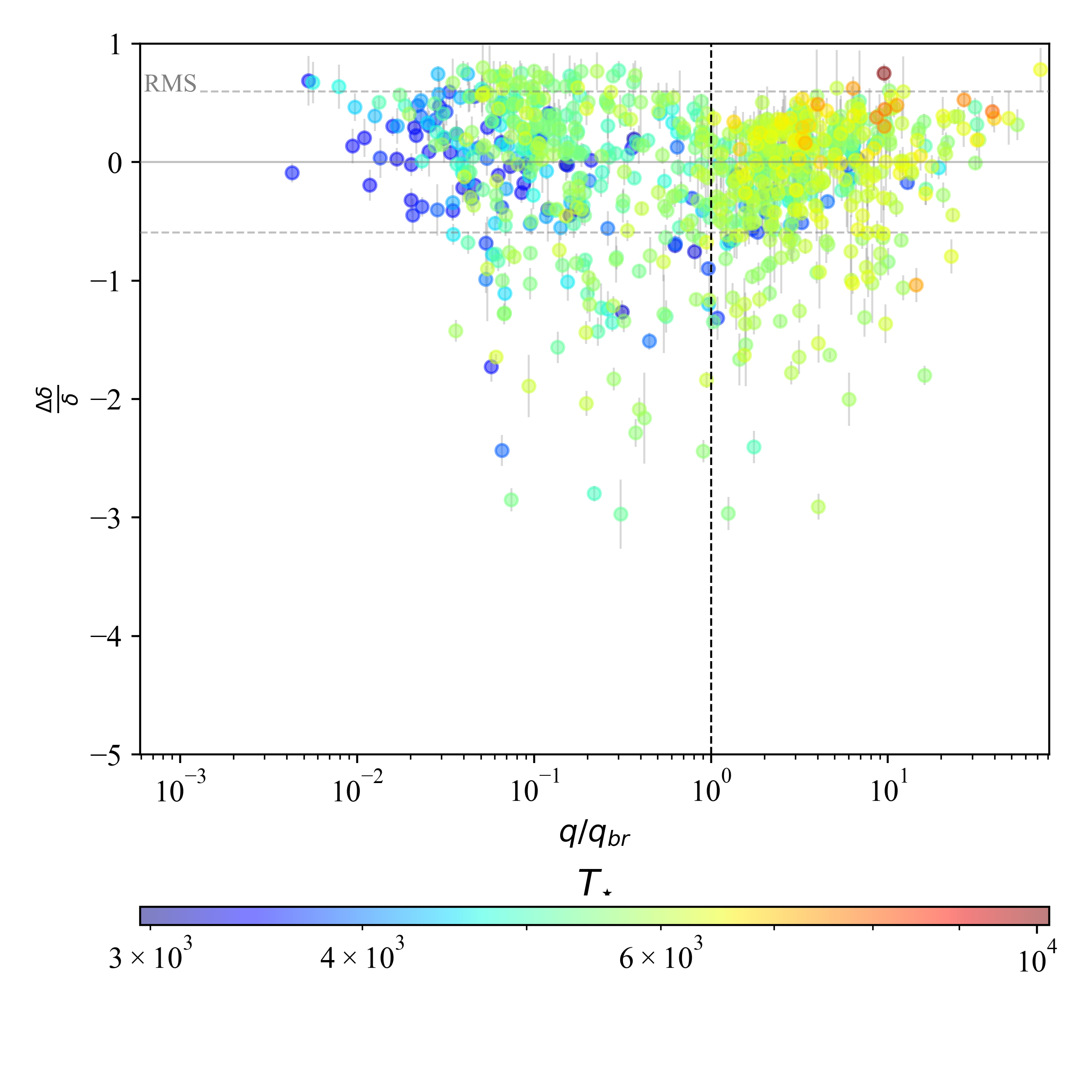}
    \includegraphics[width=0.45\textwidth]{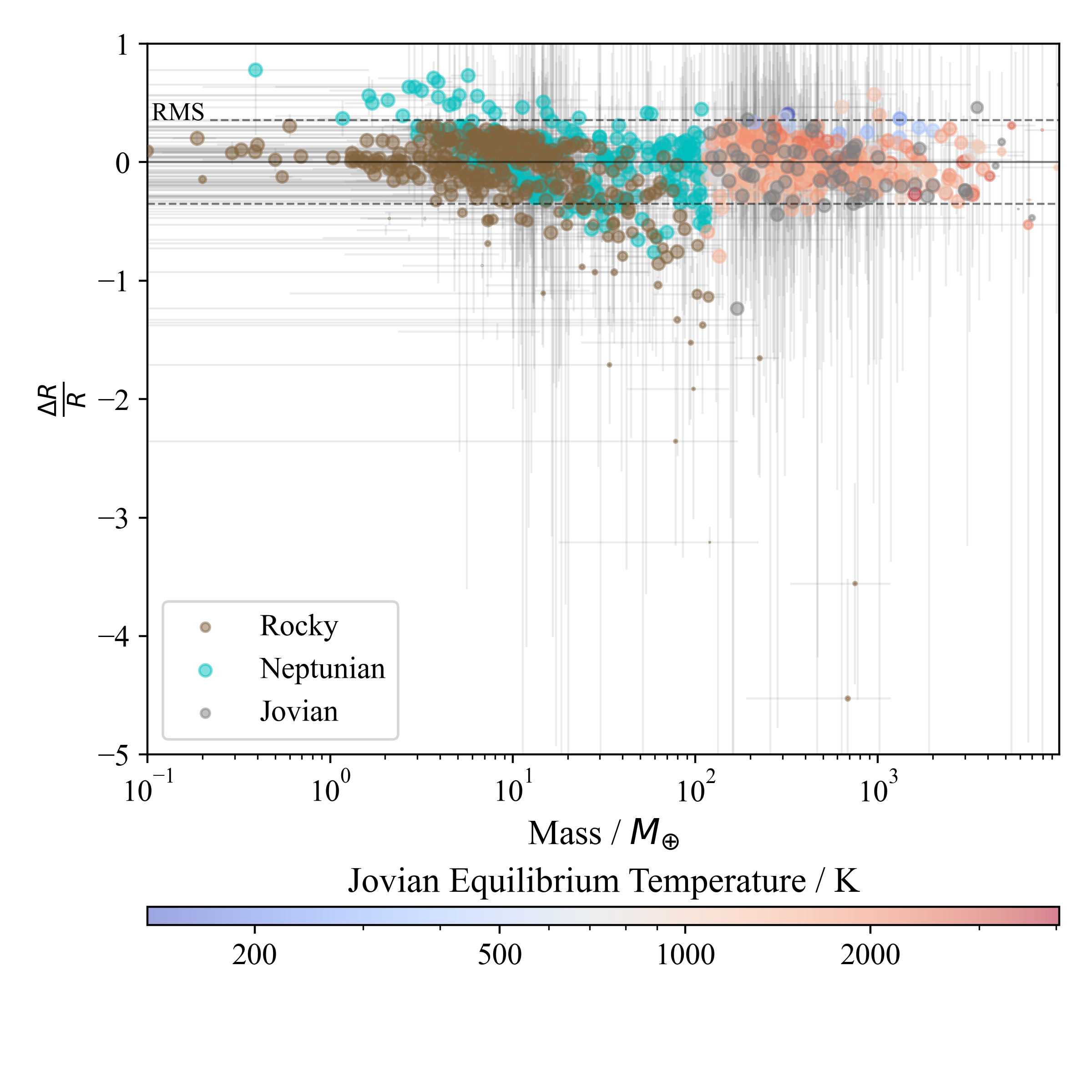}
\caption{{\em Top:} Relative residuals plotted against normalised mass ratio for the BIC-favoured PHRR model in Figure~\ref{fig:vvp_q_delta} (variable $n_1$, variable $q_{br}$, and power-law orbital period dependence). {\em Bottom:} Relative residuals from the mass--radius--temperature (MRT) relation of \citet{Edmondson2023} for their sample, colour-coded as in their work. The MRT relation shows a slightly smaller RMS scatter overall compared to the PHRR, but with more structured outlier behaviour that extends to larger negative residual values.}
\label{fig:phpp_norm_resids}
\end{figure}

The top panel of Figure~\ref{fig:phpp_norm_resids} shows the normalised residuals for this model as a function of normalised mass ratio.

There remains a group of planets hosted by cooler stars for which transit depth has been strongly over-predicted in the residuals of Figure~\ref{fig:vvp_q_delta}, though these cannot be easily distinguished from the rest of the dataset in the normalised view in Figure~\ref{fig:phpp_norm_resids}. There is no clear skewing in the residuals either, which allows the possibility that this is not a systematic effect of our model. Figure~\ref{fig:phpp_norm_resids} still demonstrates a tendency to over-predict rather than under-predict, though this may be an artefact of fitting in log space and then having to convert back to a linear scale for the residual plot.

For comparison, a similar residuals plot has been shown in the bottom panel of Figure~\ref{fig:phpp_norm_resids} for the discontinuous mass--radius--temperature (MRT) relation from \citet{Edmondson2023}. The tail at negative residuals extends further for the MRT relation than the PHRR, though most of these planets are categorised as rocky with a low fit weighting due the potential for them to be super-iron. The residuals are a little tighter for the MRT relation than for the PHRR. However, the MRT residual behaviour shows more obvious structure and extends rather further that the outlier tail for the PHRR. As previously discussed, the MRT relation can only be used for Roman planets that have either a secure mass or radius measurement, meaning the PHRR will have broader utility.

\section{Discussion}\label{sec:discussion}

\subsection{Significance of orbital period dependence}

Comparing the $\kappa$ values in Table~\ref{tab:param_results_period} reveals that the orbital period dependence is remarkably consistent across different models - transit depth decreases weakly for longer orbits. Geometrically, transit depth is approximately independent of semi-major axis due to the large distance from the observer. However, exoplanet mass--radius relations demonstrate that the composition of the planet plays a significant role in determining it's size and therefore transit depth, and the distance at which a planet forms or migrates to impacts the composition.

The distance from its host at which a planet forms determines the composition of that planet due to the temperature gradient radially outwards from the host, such that different elements condense at different distances in the protoplanetary disk. Already this impacts the bulk density of an exoplanet and thus how its size depends on mass, however subsequent migration can alter this further as different types of exoplanets respond differently to increased temperatures.

The inflating of giant planet radii, or bloating, can be attributed to the expansion of the atmosphere when heated. This is often characterised by equilibrium temperature \citep{Enoch2012} or insolation flux \citep{Thorngren2018, Sestovic2018}. In general, the radius of a giant planet around a given star should increase as the distance between them decreases, therefore increasing transit depth. In contrast, icy planets with large volatile envelopes are expected to be subject to the photoevaporation of their atmospheres if they stray too close to their host stars. This effect is even thought to be a potential formation pathway for rocky planets \citep{Lopez2013, Chen2016, Owen2017, Jin2018}, although \citet{Venturini2024} found that evaporated water in the atmospheres of volatile-rich planets acts to increase the planet radius by as much as 15\% for a $5 M_\oplus$ planet, which is a comparable effect to the bloating of giant planets.

The favoured PHRR model suggests that, planets further from their star are slightly smaller, consistent with factors such as bloating. This behaviour does agree with inflated steam envelopes around volatile-rich planets which are closer-in than their icy counterparts, but is in contrast to photoevaporation which serves to decrease the radius. It is possible that this effect is in part masked by the accompanying mass-loss. On the whole, we cannot conclude whether this relatively weak period dependence may be intrinsic to the characteristics of planet--host systems or merely a symptom of observational bias.

\subsection{Comparison to the mass ratio occurrence distribution}

\begin{figure}
\includegraphics[width=0.45\textwidth]{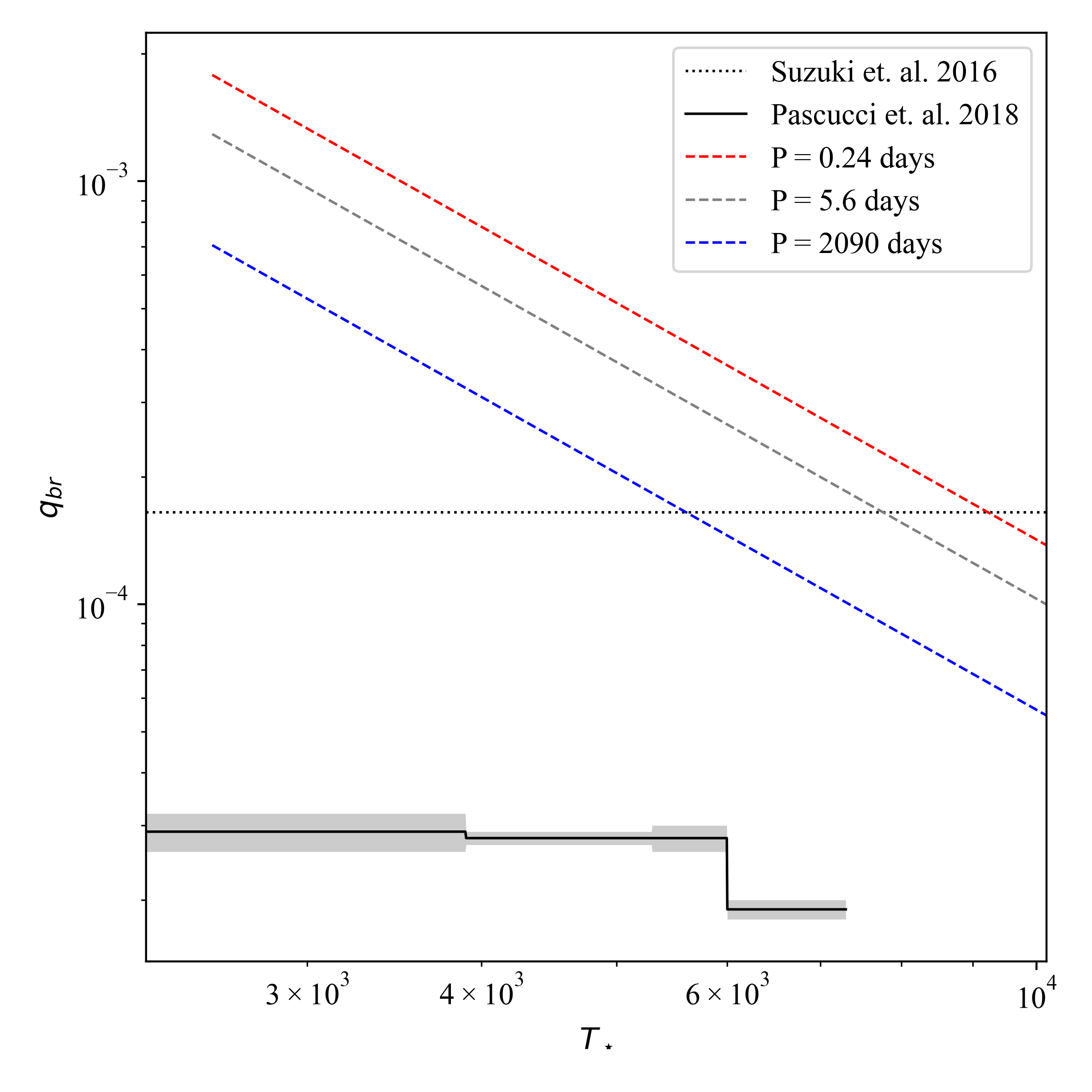}
\caption{The mass ratio break, $q_{br}$ plotted as a function of stellar effective temperature. The blue, grey and red dashed lines are for the variable $n_1$, variable $q_{br}$, period dependent PHRR from this work  for the minimum, logarithmically averaged, and maximum orbital periods in our sample, respectively. The black dotted line represents the result of $q_{br} = \num{1.65e-4}$ from \citet{Suzuki2016} for planet occurrence. The black solid line is from \citet{Pascucci2018} for planet occurrence around FGK and M stars with associated uncertainties shaded.}
\label{fig:qbr_comparison} 
\end{figure}

Given the identification in this paper of a single characteristic $q_{br}$ for the PHRR, it is worth exploring whether this is in any way related to the break in $q$ previously observed for the planet occurrence distribution. The $q_{br}$ values obtained by \citet{Suzuki2016} and \citet{Pascucci2018} for occurrence are plotted in Figure~\ref{fig:qbr_comparison} as functions of $T_\star$. The $q_{br}$ behaviour from our best PHRR (variable $n_1$ and variable $q_{br}$) is also plotted for the orbital periods spanning our dataset.

There is a coincidence of the PHRR $q_{br}$ and the occurence $q_{br}$ from \citet{Suzuki2016} for hotter stars (G-type stars and hotter). By contrast the PHRR $q_{br}$ is larger than the occurrence break value from \citet{Pascucci2018} by at least an order of magnitude for all comparable host types. However the PHRR and \cite{Pascucci2018} results both at least favour a trend of decreasing $q_{br}$ with increasing $T_\star$.

\citet{Pascucci2018} found that mass ratio break is larger for microlensing planets (longer periods) than for the transiting population (shorter periods). We find the opposite behaviour for our favoured PHRR. The period dependence ($\kappa = -0.10 \pm 0.01$) is weak but consistent with being negative across all models. Whilst our empirical relations are not corrected for sample completeness, at this point we cannot establish a link between the $q_{br}$ for the PHRR and that for planet occurrence.

\section{Conclusions}\label{sec:conclusion}

The upcoming Roman mission will detect large numbers of exoplanets using the microlensing and transit methods over similarly large Galactic distances. This haul will revolutionize our knowledge of Galactic scale exoplanet demography across both hot and cold exoplanet regimes. The primary observables of these techniques are the planet--host mass ratio (microlensing) and the planet--host radius ratio (transits) or its square, namely transit depth. Coupling together demographic information from the microlensing and transit datasets will be crucial but requires careful consideration for how to tie together the key observables. Whilst a traditional approach would be to use a planet mass--radius relation, such relations have large scatter, are based on samples that are much smaller than that which Roman will deliver, and will not allow the full range of Roman data to be used, as planet mass (or radius) will often be ill-constrained even when their respective host ratios are well determined. 

In this study we have used a carefully selected sub-sample of confirmed exoplanets from the NASA Exoplanet Archive to investigate a direct relation between planet--host mass ratio and transit depth. We refer to this as a planet--host ratio relation (PHRR).

We have shown that for a carefully selected set of 908 observed exoplanets the planet--host mass ratio and transit depth are coupled through a relatively tight power-law PHRR, especially when the stellar host effective temperature and planet orbital period are included. 

Unlike the planet mass--radius relation, the PHRR is found to be continuous over all observed scales, though it exhibits a break in power-law slope at a characteristic planet mass ratio. We have tested a number of candidate PHRR models of varying complexity and find that our most favoured model, as judged by Bayesian Information Criterion score, is one in which the transit depth scales as a two-stage power law in planet mass ratio, orbital period, and host effective temperature. Both the mass ratio break scale, and its slope below the break, are themselves found to be power-law functions of host effective temperature. Transits are predicted to be larger at fixed planet mass ratio for planets around cooler hosts. The break scale also increases for cooler stars. Our analysis does not account for detection bias and so it is not possible for us to use these relations to decouple intrinsic factors from detection bias.

We find that the power law behaviour of transit depth with orbital period is weakly inverse, but not consistent with zero, which aligns with the tendency for the atmospheres of closer-in planets to inflate via bloating or even the evaporation of water ice. But we again caution that this is population-averaged behaviour across an observationally biased dataset, and lacks the nuance to capture effects, such as photoevaporation, that can act to decrease the radii of close-in planets. Ultimately, our analysis is dominated by short-period planets and so does not have the dynamic range required to fully investigate how transit depth correlates with orbital period.

The characteristic break mass ratio for our model is larger for cooler host stars and for smaller orbital periods. This behaviour is different to that of the mass ratio break seen in some planet occurrence studies, and so we draw no obvious link between them.

The PHRR favoured in our study is purely empirical and reflects a combination of intrinsic planet--host physics and observational selection biases present in a sample that has been observed mostly through a combination of transit and radial velocity measurements. Roman will observe a far larger sample, using microlensing rather the radial velocity to measure the mass ratio of cool planets. There is therefore every reason to expect that the relevant PHRR for Roman may differ from that found here. 

Nonetheless, there are strong reasons to consider using a PHRR over a mass-radius relation to tie together the Roman transit and microlensing samples. One reason is that all Roman transit and microlensing detections will have a well determined planet-host mass or size ratio, whilst many will not have a direct planet mass or radius measurement. Use of a PHRR can therefore allow the entire sample to be used for joint demographic studies. Secondly, Roman {\em should}\/ have a significant sample of direct mass and radius measurements, as well as host temperatures, thanks to its ability to directly detect many microlensing hosts, and its ability to combine high precision relative astrometry for de-blending and false-positive discrimination, as well as multi-band photometry and even crowded-field grism spectroscopy for host characterisation. Since Roman data will also include detection efficiencies, it should be possible from Roman data alone to self-calibrate the appropriate PHRR from sub-samples with secure host mass or radius measurements. Whilst the mass and radius measurements will not come from the same planets, parameters for a universal PHRR like the one in this paper can either be retrieved along with demographic model parameters, or be shown to be in tension with a convergent joint demographic analysis.

\section*{Acknowledgments}

We thank the anonymous referee for their comments which improved the paper. Kathryn Edmondson is funded by the Science and Technology Facilities Council (STFC) via a PhD studentship. This research has made use of the NASA Exoplanet Archive, which is operated by the California Institute of Technology, under contract with the National Aeronautics and Space Administration under the Exoplanet Exploration Program. The authors acknowledge the use of \texttt{Astropy} \citep{Astropy2013, Astropy2018}, \texttt{Matplotlib} \citep{Hunter2007}, \texttt{NumPy} \citep{vanderWalt2011}, \texttt{pandas} \citep{McKinney2010, pandas2020}, and \texttt{SciPy} \citep{Virtanen2020}.

\section*{Data Availability}

The Planetary Composite Parameters dataset from the NASA Exoplanet Archive used for this work can be accessed here: \url{https://exoplanetarchive.ipac.caltech.edu/cgi-bin/TblView/nph-tblView?app=ExoTbls&config=PSCompPars}



\bibliographystyle{mnras}
\bibliography{references} 








\bsp	
\label{lastpage}
\end{document}